\documentclass[aps, showpacs, showkeys,nofootinbib,floatfix]{revtex4}

\usepackage{amssymb}
\usepackage{amsmath}
\usepackage{graphicx}
\usepackage{hyperref}

\begin{document}

\title{The Influence of Free Quintessence on Gravitational Frequency Shift and Deflection of Light with 4D momentum}

\author{Molin Liu}
\email{mlliudl@student.dlut.edu.cn}
\author{Jianbo Lu}
\email{lvjianbo819@163.com}
\author{Yuanxing Gui}
\email{guiyx@dlut.edu.cn}

\affiliation{School of Physics and Optoelectronic Technology,
Dalian University of Technology, Dalian, 116024, P. R. China}

\begin{abstract}
Based on the 4D momentum, the influence of quintessence on the
gravitational frequency shift and the deflection of light are
examined in modified Schwarzschild space. We find that the frequency
of photon depends on the state parameter of quintessence $w_q$: the
frequency increases for $-1<w_q<-1/3$ and decreases for
$-1/3<w_q<0$. Meanwhile, we adopt an integral power number $a$ ($a =
3\omega_q + 2$) to solve the orbital equation of photon. The
photon's potentials become higher with the decrease of $\omega_q$.
The behavior of bending light depends on the state parameter
$\omega_q$ sensitively. In particular, for the case of $\omega_q =
-1$, there is no influence on the deflection of light by
quintessence. Else, according to the H-masers of GP-A redshift
experiment and the long-baseline interferometry, the constraints on
the quintessence field in Solar system are presented here.
\end{abstract}



\maketitle

\section{Introduction}
\label{intro} It is well known that the universe accelerating is
proved by the type Ia Supernovae (SNe Ia) \cite{Ia} and the cosmic
microwave background (CMB) \cite{CMB} in Wilkinson Microwave
Anisotropy Probe (WMAP). The large scale distribution of galaxies
\cite{largescalar} shows the existence of cold dark matter (CDM)
with the ratio of $\Omega_{CDM} = 0.27 \pm 0.04$. The current
universe is dominated by the dark energy component which contributes
$\Omega_{de} = 0.67 \pm 0.06$ to the critical energy density. Else,
dark energy should also have a negative pressure to ensure the
acceleration of universe.

These astronomical observations constrain current state parameter
$\omega$ to be close to the cosmological constant case
\cite{cosconstant} which corresponds to a fluid with $\omega = - 1$.
Actually, the observations also indicate a little bit time evolution
of $w$. Thus, many people considered the corresponding situation in
which the state equation of dark energy changes with time. From
particle physics, the scalar fields is natural candidates of dark
energy. A wide variety of scalar field dark energy models have been
proposed such as quintessence \cite{q}, K-essence \cite{k-essence},
tachyon field \cite{Tachyon}, Phantom (ghost) \cite{phantom},
dilatonic dark energy \cite{dilatonic} and so on. The quintessence
model refers to a minimally coupled scalar field with a potential
which decreases as the field increases. The field is coupled by the
gravity through a Lagrangian taking the form
\begin{equation}\label{lagrangian}
    L_{q} = \sqrt{g}\left(\frac{1}{2} \partial \phi^2 -
    V(\phi)\right),
\end{equation}
where $\phi$ is the scalar field and $V(\phi)$ is the potential.

On the other hand, many people have discussed minutely how to
combine the quintessence energy momentum tensor with Einstein
equations \cite{Chernin} \cite{Gonzalez} \cite{Kiselev}. However,
the initial solutions obtained by \cite{Chernin} \cite{Gonzalez}
have no horizon and no `hair'. Hence, the black hole can not be
formed in those metrics. Later, Kiselev \cite{Kiselev} adopted a
nonzero off-diagonal energy momentum tensor being proportional to
diagonal terms, i.e. $C(r)\neq 0 \propto B(r)$, where $C(r)$ and
$B(r)$ are the coefficients of the energy momentum tensors. The
energy momentum tensor of the quintessence $T_{\mu}^{\nu}$ are
\begin{equation}\label{momentumquintess}
    T_{t}^{t} = A(r),\ \ \ T_{t}^{j} = 0, \ \ \ T_{i}^{j} =
    C(r)r_{i}^{j} + B(r) \delta _{i}^{j}.
\end{equation}
The constant coefficient $C(r)/B(r)$ satisfies the additivity and
linearity conditions. So a exact black hole solution surrounding by
quintessence matter
 is obtained by the static coordinates \cite{Kiselev}. Since then, many people have performed its quasinormal modes \cite{quasinormal},
entropy \cite{entropy}, geodesic precession\cite{geodetic} and so
on. However, as far as we know, there is no work discussing its
influence on gravity test in the view of 4D momentum. In this paper,
we start from the 4D momentum and calculate carefully the
gravitational shift and the deflection of light in the modified
Schwarzschild space encoded the quintessence matter. From the
consideration of the supernovae dimming, we should make the dark
energy satisfies $\omega \leq -2/3$ to fit the observations
\cite{Efstathiou}. But in this paper we evaluate the quintessence
matter with the state parameter in the range of $\omega_q \in [-1,
0]$. Meanwhile, we also compare the result of quintessence with that
of cosmological constant \cite{SdS} to verify its rationality based
on the fact of that $\Lambda$CDM is reduced from a special
quintessence field of $\omega_q = -1$ to a certain extent. It should
also be noted that the problems of gravitational frequency shift and
light deflection are studied in the context of central gravitational
field rather than the cosmology here.

This paper is organized as follows: in section II, we present the
Kiselev solution, i.e. the Schwarzschild black hole surrounded by
quintessence matter. In section III, we calculate the frequency
gravitational shift of photons in which the frequency is inverse
proportional to local $\sqrt{g_{00}}$. In section IV, we study the
deflection of light according to the different state parameter
$\omega_q$. In section V, we use the astronomical observations to
constrain the quintessence and examine whether the observable effect
is big enough to be measured. Section VI is a conclusion. We adopt
the signature $(+, -, -, -)$ and put $\hbar$, $c$, and $G$ equal to
unity.
\section{Schwarzschild space surrounded by static spherically symmetric quintessence matter}
Before performing the gravitational test of modified general
relativity (GR), we would like to introduce Kiselev's black hole
solution \cite{Kiselev}. The derivation starts from a spherically
symmetric static gravitational field,
\begin{equation}\label{ssmetric}
    d s^2 = e^{\nu (r)} dt^2 - e^{\lambda (r)} dr^2 - r^2 (d\theta^2 + \sin^2\theta
    d\varphi^2).
\end{equation}
It assumes that the quintessence matter distributes evenly outside
black hole and the energy momentum tensor takes the forms
\begin{eqnarray}
  T_{t}^{t} &=& \rho_q (r), \\
  T_{i}^{j} &=& \rho_q (r) \gamma \left[-(1 + 3B)\frac{r_{i}r^{j}}{r_{n}r^{n}} + B
  \delta_{i}^{j}\right],
\end{eqnarray}
where $\rho_q$ is the density of quintessence matter. The parameter
$B$ depends on the internal structure of quintessence. The isotropic
average over the angle components is
\begin{equation}\label{average}
    <T_{i}^{j}> = -\rho_{q}(r)\frac{\gamma}{3}\delta_{i}^{j} =
    -p_{q}(r) \delta_{i}^{j},
\end{equation}
where the relationship $<r_{i}r^{j}> =
\frac{1}{3}\delta_{i}^{j}r_{n}r^{n}$ is used and the state equation
is $P_{q} = \omega_q \rho_q$ where $\omega_q = \gamma/3$.

According to the additivity and linearity principle, i.e. $T_{t}^{t}
= T_{r}^{r}$, the free parameter $B(\omega_q)$ can be fixed as
\begin{equation}\label{Bomega}
    B = - \frac{3\omega_q + 1}{6\omega_q}.
\end{equation}
Meanwhile, this assumption directly leads to a relationship $\lambda
+ \nu = 0$. Hence the energy momentum tensor can be written as
follows,
\begin{eqnarray}
  T_t^t &=& T_r^r = \rho_q \label{add1},\\
  T_{\theta}^{\theta} &=& T_{\varphi}^{\varphi} =
  -\frac{1}{2}\rho_q (3\omega_q + 1). \label{theta}
\end{eqnarray}
If we set $\lambda = -\ln (1 + f)$, a differential equation of $f$
can be obtained
\begin{equation}\label{fequation}
   r^2\frac{d^2 f}{d r^2} + 3 (1 + \omega_q) r \frac{d f}{d r} + (3\omega_q + 1) f=
   0,
\end{equation}
which has two exact solutions
\begin{eqnarray}
  f_g &=& \frac{\alpha}{r^{3\omega_q + 1}} \label{fg}, \\
  f_{BH} &=& -\frac{r_g}{r}\label{fbh},
\end{eqnarray}
where $\alpha$ and $r_g$ are the normalization factors. $f_{BH}$ is
the ordinary point-like black hole solution such as the
Schwarzschild black hole. The quintessence density is
\begin{equation}\label{rho}
 \rho_q =
\frac{\alpha}{2}\frac{3\omega_q}{r^{3(1+\omega_q)}}.
\end{equation}
So if we require the density of energy positive, $\rho_q > 0$, we
deduce that $\alpha$ is negative for $\omega_q$ negative. The
curvature has the form of
\begin{equation}\label{curvature}
    R = 2 T_{\mu}^{\mu} = 3 \alpha \omega_q \frac{1-3\omega_q}{r^{3(\omega_q +
    1)}}.
\end{equation}
Apparently, r = 0 is the singularity for $\omega_q\ \neq \{-1,\ 0,\
1/3 \}$. Combining linearly solutions (\ref{fg}) and (\ref{fbh}), we
can get a general solution.
\begin{equation}\label{e}
    e^{-\lambda} = 1 - \frac{r_g}{r} - \sum_{n}\left(\frac{r_n}{r}\right)^{3\omega_n +
    1},
\end{equation}
where $n$ indicates several fields and $r_q$ is a normalization
factor which has the dimension of length. The free quintessence
creates an outer horizon of de Sitter universe at $r = r_q$ for
\begin{equation}\label{outerhorizon}
    -1 < \omega_q < -\frac{1}{3},
\end{equation}
and also generates an inner horizon of black hole at $r = r_q$ for
\begin{equation}\label{innerhorizon}
    -\frac{1}{3} < \omega_q <0.
\end{equation}
 Here we only consider the influence of quintessence matter on
Schwarzschild space. The metric of Schwarzschild space is modified
by a new form,
\begin{eqnarray}\label{metricq}
d s^2 = \bigg{(}1 - \frac{2M}{r} - \frac{\alpha}{r^{3\omega_q +
    1}}\bigg{)} d t^2 - \bigg{(}1 - \frac{2M}{r} - \frac{\alpha}{r^{3\omega_q +
    1}}\bigg{)}^{-1} d r^2 -r^2 (d \theta^2 + \sin^2\theta d
    \varphi^2),
\end{eqnarray}
where $M$ is the black hole mass. When $\alpha = 0$ this metric
reduces to a pure Schwarzschild case. Otherwise, when $\omega_q =
-1$ this metric reduces to a Schwarzschild-de Sitter one. It should
be noticed that comparing with the large total mass of dark energy,
the relatively small mass of black hole can be treated as a
constant.
\section{Frequency Gravitational Shift}
When photon propagates in a stable gravitational field, the
stationary observers at different positions obtain different
frequencies. This phenomenology is the so called gravitational
redshift which is one of the special effects purely due to curved
space. Certainly, it is also a natural result from the equivalence
principle in general relativity. For the static metric $g_{\mu\nu}$
(\ref{metricq}), the photon's energy can be written as
\begin{equation}\label{energy}
    E = p_{\mu}U^{\mu},
\end{equation}
where $p_{\mu}$ is the 4D momentum of photon, $U^{\mu}$ is the 4D
velocity of observer which is in the form
\begin{equation}\label{velocity}
    U^{\mu} = \frac{1}{\sqrt{g_{00}}}(1,\ 0,\ 0,\ 0).
\end{equation}
Submitting Eq. (\ref{velocity}) into Eq. (\ref{energy}), the energy
$E$ can be rewritten as
\begin{equation}\label{energy1}
    E = \frac{ p_{0}}{\sqrt{g_{00}}}.
\end{equation}
Using the Planck relation $E = h \nu$, the frequency and metric
component $g_{00}$ yield
\begin{equation}\label{frequency}
    \sqrt{g_{00}}\nu = p_{0}/h.
\end{equation}
The right hand side of Eq. (\ref{frequency}) is a constant since
$p_0$ is a conserved parameter when photon moves in this stable
space.

When Schwarzschild space is surrounded by the free static
spherically symmetric quintessence matter, the time component of
metric (\ref{metricq}) is
\begin{equation}\label{g00}
    g_{00} = 1- \frac{2M}{r} - \frac{\alpha}{r^{3\omega_q + 1}}.
\end{equation}
Meanwhile, we assume that the emitter is at $x_0 = (r_0,\ 0,\ 0,\
0)$ and the receiver is at $x = (r,\ 0,\ 0,\ 0)$. The photon signal
has fixed spatial coordinates and their 4-velocity is tangent to the
static Killing field $\xi^a$. Furthermore, the photon frequency is
inverse proportion to the local $\sqrt{g_{00}}$ and the frequencies
of emitter and receiver must satisfy
\begin{equation}\label{frequency2}
    \frac{\nu}{\nu_0} =
    \sqrt{\frac{g_{00}(r_0)}{g_{00}(r)}}.
\end{equation}
Submitting the lapse function Eq. (\ref{g00}) into above ratio $\nu
/ \nu_0$, we can obtain
\begin{eqnarray}\label{f3}
\frac{\nu}{\nu_0} =
    \sqrt{\frac{g^{(0)}_{00}(r_0)}{g^{(0)}_{00}(r)}}\Bigg{\{}1 + \frac{\alpha}{2}\Bigg{(}\frac{1}{g^{(0)}_{00}(r)r^{3\omega_q +1}} - \frac{1}{g^{(0)}_{00}(r_0) r_0^{3\omega_q +
    1}}\Bigg{)}\Bigg{\}},
\end{eqnarray}
where $g^{(0)}_{00} = 1 - 2M/r$ and we assume $2M/r \gg
\alpha/r^{3\omega_q + 1}$. In the weak field limit
\begin{equation}\label{weaklimit}
    M/r \ll 1,
\end{equation}
Eq. (\ref{f3}) can be simplified to
\begin{equation}\label{f4}
\frac{\nu}{\nu_0} = 1 + \frac{M}{r} - \frac{M}{r_0} + \Delta_{\nu},
\end{equation}
where
\begin{equation}\label{Delta}
    \Delta_{\nu} = \frac{\alpha}{2}\left(\frac{1}{r^{3\omega_q + 1}} -
\frac{1}{r_0^{3\omega_q + 1}}\right).
\end{equation}
Hence, the relative gravitational redshift (or redshift parameter)
$z$ is
\begin{equation}\label{z}
    z = \frac{\nu - \nu_0}{\nu_0} = \frac{M}{r} - \frac{M}{r_0} +
\Delta_{\nu},
\end{equation}
where the first and second terms are the result of GR and the last
one $\Delta_{\nu}$ describes the amendment to Schwarzschild space by
quintessence matter. Furthermore, the modified term $\Delta_{\nu}$
satisfies the following relation
\begin{equation}\label{Deltarelation}
    \Delta_{\nu} \longrightarrow  \left\{
    \begin{array}{c}
   \ \  > 0 \ \ \ \ for\ \  -1 < \omega_q < -1/3,\\
    < 0 \ \ \ \ for\ \  -1/3< \omega_q < 0.
    \end{array}
    \right.
\end{equation}
Comparing with the usual pure Schwarzschild result \cite{book}, we
can find the redishift increases for $-1 < \omega_q < -1/3$ and
decreases for $-1/3 < \omega_q < 0$.

\section{Deflection of Light}
\begin{figure*}
\begin{center}
\resizebox{0.6\textwidth}{!}{%
  \includegraphics{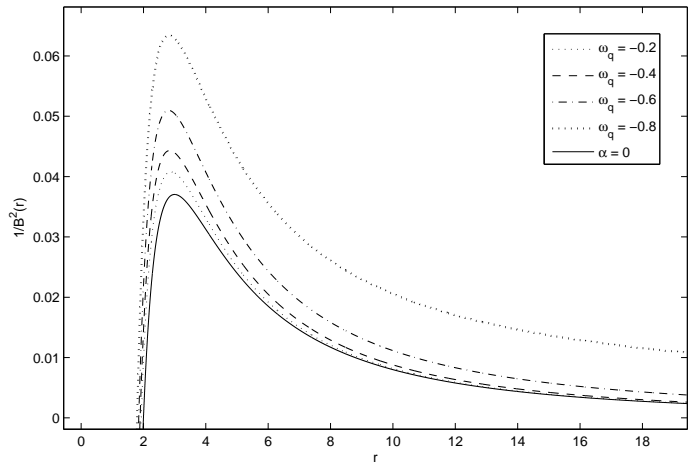}
} \caption{The effective potentials $1/B^2 (r)$ of photons with
$\alpha = -0.05$ and unit mass $M = 1$. The corresponding
Schwarzschild case $\alpha = 0$ is drawn by solid line.} \label{f1}
\end{center}
\end{figure*}

Because that the movement of photon is described by the null line
element $ds = 0$,  $s$ can not be treated as the affine parameter
any more. The original definition of 4D momentum $p^{\mu} = m
\frac{dx^{\mu}}{d\tau}$ is not suitable to describe massless
photons. Considering the above reasons, we choose an arbitrary
scalar parameter $\zeta$ as a new affine parameter. Therefore the 4D
momentum of photon is redefined as
\begin{equation}\label{momentum}
    p^{\mu} = \frac{d x^{\mu}}{d \zeta}.
\end{equation}
When photon goes in the static spherically symmetric space
(\ref{metricq}), $p_0$ and $p_3$ are still conserved parameters. The
first and second motion equations about $t$ and $\varphi$ are
obtained naturally in the following equations:
\begin{eqnarray}
   && r^2 \frac{d\varphi}{d \zeta} \label{m1} = L,\\
   && (1 - \frac{2M}{r} - \frac{\alpha}{r^{3\omega_q + 1}}) \frac{d t}{d
   \zeta} = E\label{m2}.
\end{eqnarray}
The third motion equation can be given by the normalization relation
of photons $g_{\mu\nu} p^{\mu}p^{\nu}$ = 0,
\begin{equation}\label{m3}
    \left(\frac{d r}{d \zeta}\right)^2 = E^2 - \frac{L^2}{r^2}\left(1 - \frac{2M}{r} - \frac{\alpha}{r^{3\omega_q +
    1}}\right).
\end{equation}
The dynamic evolution of photon is determined fully by
Eqs.(\ref{m1}), (\ref{m2}) and (\ref{m3}) in this space. Combining
Eq. (\ref{m1}) with (\ref{m3}), the photon's trajectory equation can
be gotten as
\begin{equation}\label{trajectoryequation}
    \left(\frac{1}{r^2}\frac{d r}{d \varphi}\right)^2 =
    \left(\frac{E}{L}\right)^2 - \frac{1}{r^2} \left( 1- \frac{2M}{r} - \frac{\alpha}{r^{3 \omega_q +
    1}}\right).
\end{equation}
Meanwhile, we define two new parameters, one is the impact parameter
$b = L/E$, which means the effective sighting range, the other is
the photon's effective potential $1/B^2(r)$ where
\begin{equation}\label{ep}
    B (r) = r \left(1 - \frac{2M}{r} - \frac{\alpha}{r^{3 \omega_q + 1}}
    \right)^{-1/2}.
\end{equation}
The effective potential, which is illustrated by
\textbf{Fig.}\ref{f1}, becomes higher with smaller state parameter
$\omega_q$.

According to Eq. (\ref{ep}), photons' orbital equation
(\ref{trajectoryequation}) can be rewritten as
\begin{equation}\label{or3}
    \left(\frac{1}{r^2} \frac{d r}{d \varphi}\right)^2 =
    \frac{1}{b^2} - \frac{1}{B^2(r)}.
\end{equation}

Calculating the first order derivation of the trajectory equation
(\ref{trajectoryequation}) with respect to $\varphi$, the photon
Binet equation can be obtained as follows,
\begin{equation}\label{traqu2}
    \frac{d^2 u}{d \varphi^2} = -u + 3 u^2 + \frac{3\alpha (\omega_q + 1)}{2 M^{3\omega_q + 1}} u^a,
\end{equation}
where $u = M/r$ and $a = 3 \omega_q +2$. In the right hand side of
the above equation, the second term is the general relativity
correction and the third one is the quintessence contribution.
Because $u$ can be treated as a small quantity in the weak field
regime, we can solve Eq. (\ref{traqu2}) by the successive
approximation method.

The negative pressure of quintessence matter ($p_q = \omega_q
\rho_q$) supplies the state parameter $-1\leq \omega_q \leq 0$. The
exponential $a$ in the additional term of Eq. (\ref{traqu2}) is in
the scope $-1\leq a \leq 2$. It is more difficult to solve
Eq.(\ref{traqu2}) just by the undetermined state parameter
$\omega_q$ and normalization $\alpha$. In order to solve this
problem, we select the solvable integers ($a = -1,\ 0,\ 1,\ 2$) in
the range of $\omega_q \in [-1, 0]$. The four cases' results, which
are $(a =2, \omega_q = 0)$, $(a =1, \omega_q = -1/3)$, $(a = 0,
\omega_q = -2/3)$ and $(a = -1, \omega_q = -1)$ respectively, are
listed in the following \textbf{TABLE \ref{table1}}. Here, the
parameters $u_0$, $y$ and $y_{0}$ are $u_0 = \frac{GM}{R}$, $y = u +
\frac{\sqrt{1-6\alpha M} - 1}{6}$ and $y_0 = u_0 + \frac{\sqrt{1 - 6
\alpha M} - 1}{6}$, where $R$ is the solar radius. The calculation
detail are shown in Appendix \ref{appendixa}, \ref{appendixb},
\ref{appendixc} and \ref{appendixd}.

\begin{figure*}
\begin{center}
\resizebox{0.5\textwidth}{!}{%
  \includegraphics{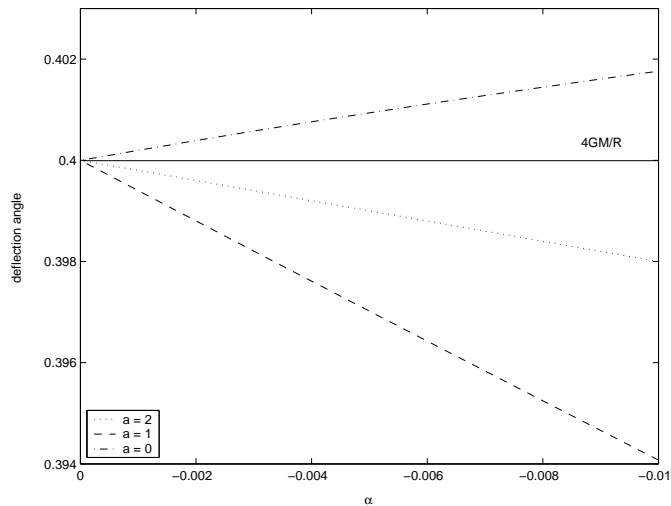}
} \caption{Deflection angle $\delta\beta$ versus quintessence's
normalization parameter $\alpha$ with $M = 1$, $R = 10$. The solid
line denotes the pure Schwarzschild case \cite{book} whose
deflection angle $\delta\beta = 4GM/R$ corresponds to the case of $a
= -1$.} \label{fig2}
\end{center}
\end{figure*}

The forms of influence on light deflection by quintessence are
different according to various state parameters $\omega_q$. The
deflection angles depend sensitively on quintessence's normalization
parameter $\alpha$. The deflection angles are illustrated in
\textbf{Fig.} \ref{fig2}. For the case $\omega_q = -2/3$, the
deflection angle increases monotonically as $| \alpha |$ increases.
On the contrary, the deflection angles decrease with increasing $|
\alpha |$ in the cases of $\omega_q = -1/3$ and $\omega_q = -2/3$.
Meanwhile, comparing to the pure Schwarzschild case, the deflection
becomes larger in the case of $a = 0$ and becomes smaller in the
cases $a = 1$ and $2$. These variations are caused by the fact that
the modified term $\frac{3\alpha (\omega_q + 1)}{2 M^{3\omega_q +
1}} u^a$, in the photon's orbital equation (\ref{traqu2}), gives the
different solutions. The behaviors of deflection heavily depend on
the quantities of state parameter $\omega_q$. It should be notice
that in order to easily analyse the results, we recover the
international system of units from the natural unit.

\begin{table*}[!h]\scriptsize
\caption{The influence of quintessence on deflection of light}
\label{table1}
\begin{tabular}{llll}
     \hline\noalign{\smallskip} state parameter $\omega_q$ & reduced Binet equation& the first order approximate solution & deflection angle \\
     \noalign{\smallskip}\hline\noalign{\smallskip}
     $a =2,\omega_q =0$&$\frac{d^2 u}{d \varphi^2} + u = \left(3 + \frac{3\alpha}{2M}\right) u^2$&$u = u_0 \cos\varphi + \left(1 + \frac{\alpha}{2M}\right)u_0^2 (1 +
    \sin^2\varphi)$&$\delta\varphi =2\beta = 4\left(1 + \frac{\alpha}{2 M}\right)\frac{GM}{R}$\\
     $a =1, \omega_q = -1/3$&$\frac{d^2 u}{d \varphi^2} + (1 - \alpha) u = 3 u^2$&$u = u_0 \cos(\varphi\sqrt{1 - \alpha}) + \frac{u_0^2}{1 - \alpha}\left(1 + \sin^2(\varphi\sqrt{1 - \alpha})\right)$&$\delta\varphi = \frac{4}{(1-\alpha)^{3/2}}\frac{GM}{R}$\\
     $a =0, \omega_q = -2/3$&$\frac{d^2 u}{d\varphi^2} + u - \frac{\alpha M}{2} = 3 u^2$&$y = y_0 \cos [\varphi(1 - 6 \alpha M)^{1/4}] + \frac{y_0^2}{\sqrt{1 - 6 \alpha M}} \{1 + \sin^2 [ \varphi (1 - 6 \alpha M)^{1/4}] \}$&$\delta\varphi = \frac{4 y_0}{(1 - 6 \alpha M)^{3/4}}$\\
     $a =-1,\omega_q = -1$&$\frac{d^2 u}{d\varphi^2} + u  = 3 u^2$&$u = u_0 \cos\varphi + u_0^2 (1 +\sin^2\varphi)$&$\delta\varphi = 4\frac{GM}{R} $\\
     \noalign{\smallskip}\hline
\end{tabular}
\end{table*}

\section{experimental constraints}
We know that astronomical observations, such as the gravitational
lensing measurements, become important in determining cosmological
parameters. Here we will do this work in the view of black hole. We
compare the theoretical predictions of gravitational frequency shift
and deflection of light with the typical experiments and give the
experimental constraints on the cosmological parameters.

Firstly, we consider the case of gravitational frequency shift. The
results of frequency comparison/clock comparison are given by the
hydrogen masers frequency standard in the GP-A redshift experiment
\cite{Vessot}. The continuous microwave signals are generated from
H-masers located in the spacecraft and at an Earth station where the
spacecraft was launched nearly vertically upward to 10000 km. This
experiment reached a $10^{-14}$ accuracy. So, if we consider the
quintessence field's effect, the parameter $\alpha$ has to satisfy
the constraint
\begin{equation}\label{constraintalpha}
    |\alpha| \lesssim 2 \times 10^{-14} \times
    \frac{(r r_0)^{3\omega_q +1}}{|r^{3\omega_q + 1}-r_0^{3\omega_q+1}|}.
\end{equation}

The constraints of field parameter $\alpha$ are presented in
\textbf{TABLE \ref{table2}}. Generally speaking, the upper limit of
field parameter $|\alpha|$ increases with bigger state parameter
$\omega_q$. When $\omega_q$ crosses over $-1$ the maximum $|\alpha|$
is below the value of $10^{-28}$, which is consistent with the
cosmological constant case $|\alpha| \lesssim 10^{-28}$ \cite{SdS}.
However, there is a singular point located at $\omega = -1/3$ since
the definition of $\omega_q = -1/3$ does not exist in
Eq.(\ref{constraintalpha}). Else, when $\omega_q \rightarrow -1/3$,
i.e. $\Delta_{\nu} \rightarrow 0$, the redshift parameter $z$
(\ref{z}) reduces to the result of pure Schwarzschild case.

\begin{table*}[!h]\scriptsize
\caption{The constraint on field parameter $\alpha$ by H-masers of
GP-A redshift experiment.} \label{table2}
\begin{center}
\begin{tabular}{lllllll}
     \hline\noalign{\smallskip}
           $\omega_q$ & $0$ & $-0.2$ &$-0.4$ &$-2/3$ &$-0.8$ &$-1$ \\
     \noalign{\smallskip}\hline\noalign{\smallskip}
    Estimate on $\alpha$ &\ \ $|\alpha|\lesssim 3 \times 10^{-7}$&\ \ $|\alpha|\lesssim 6 \times 10^{-11}$&\ \ $|\alpha|\lesssim 8\times 10^{-15}$&\ \ $|\alpha|\lesssim 5 \times 10^{-21}$&\ \ $|\alpha|\lesssim 6 \times 10^{-24}$&\ \ $|\alpha|\lesssim 3 \times 10^{-28}$\\
\noalign{\smallskip}\hline
\end{tabular}
\end{center}
\end{table*}

Secondly, we consider the other case of deflection of light. In
order to obtain experimental constraints from the light deflection
result, the final light deflection angle $\delta \varphi$ should be
expressed in terms of the deviation $\Delta_{LD}$ from the general
relativity prediction $\delta \varphi _{GR}$ for the Sun.
\begin{equation}\label{deviation}
\delta \varphi = \delta \varphi_{GR}(1+\Delta_{LD}),
\end{equation}
where $\delta \varphi_{GR} = 4 G M /R$. The best available
constraints on $\delta \varphi _{GR}$ come from long-baseline radio
interferometry which shows that $|\Delta_{LD}| \leq 0.0017$
\cite{Robertson}. Submitting the deflection angles of $\omega_q =
-1,\ -2/3,\ -1/3,\ 0$ shown in \textbf{TABLE \ref{table1}} into
Eq.(\ref{deviation}), we can obtain the corresponding constraints on
the field parameter $\alpha$, which are present in \textbf{TABLE
\ref{table3}.} The constraint on $\alpha$ becomes stronger with the
decreasing state parameter $\omega_q$.
\begin{table*}[!h]
\caption{Estimates on $\alpha$ from Solar system observation.}
\label{table3}
\begin{center}
\begin{tabular}{lcl}
     \hline\noalign{\smallskip}
           $\omega_q$ & $\Delta_{LD}$ & Estimate on $\alpha$ \\
     \noalign{\smallskip}\hline\noalign{\smallskip}
    0 &$\frac{\alpha}{2M}$&$|\alpha| \lesssim 10^{27}$\\
    $-1/3$ &$1-(1-\alpha)^{-3/2}$&$|\alpha| \lesssim 10^{-3}$\\
    $-2/3$&$(1+\frac{9}{2}\alpha M)(1-\frac{\alpha R}{2G})-1$&$|\alpha| \lesssim 10^{-34}$\\
    $-1$&0&---\\
\noalign{\smallskip}\hline
\end{tabular}
\end{center}
\end{table*}

\section{Conclusion}
In this paper, we have studied the influence of free quin-tessence
on gravitational frequency shift and deflection of light based on
the 4D momentum. We summarize what have been achieved.

1. The influence of quintessence matter on our space is expressed
mathematically as a new metric form. Taking advantage of the
symmetries of Kiselev solution, we can avoided to solve directly the
original geodesic equation. The inner product $u^a \xi_a$ is
constant along the geodesic in a Killing field $\xi^a$ with a
geodesic tangent $u^a$. In this paper, starting from the 4D momentum
the geodesics problem is reduced to the problem of one-dimensional
motion of a particle in an effective potential. The null geodesic of
the modified Schwarzschild geometry containing the quintessence
matter is solved by the usual method \cite{book}. We analyze the
behavior of light ray in the weak field regime $r \gg M$. Also, this
result can be applied to the exterior field of ordinary body such as
the Sun or more large astronomical scale. Meanwhile, the
quintessence matter is also treated as a very small component in
whole space, i.e. $M/r \gg \alpha/r^{3\omega_q + 1}$. These
appropriate approximation can help us to simplify the calculation of
the gravitational redshift and the bending of light.

2. The small $g_{00}$ in a metric is corresponding to the high
frequency because the frequency of photon is inverse proportional to
local $\sqrt{g_{00}}$, i.e. $\nu \propto 1/\sqrt{g_{00}}$. The
gravitational redshift expression (\ref{z}) is obtained naturally in
the weak field limit. It is clearly that the frequency is larger (or
smaller) for $ -1 < \omega_q < -1/3$ (or $-1/3 < \omega_q < 0$) than
that of the pure Schwarzschild case from the result of redshif
(\ref{f4}).

3. Comparing with the gravitational redshift, the deflection of
light is more or less complex with the uncertain power $a$ in the
orbital equation (\ref{traqu2}). We choose the solvable integral
numbers $a$ in the range of $\omega_q \in [-1, 0]$ to obtain the
deflected angle. The solutions of the orbital equation
(\ref{traqu2}) depend sensitively on the value of $\omega_q$.
Furthermore, when $a = -1$, i.e. $\omega_q = -1$, there is no
influence on bending of light by quintessence matter, which is in
accordance with the theoretical prediction of SdS case \cite{SdS}.
This point also illustrates well that the border case of the
extraordinary quintessence $\omega_q = -1$ covers the cosmological
constant term \cite{cosconstant}.

4. The parameter $\alpha$ is of course a crucial parameter, which
indicates the influence of quintessence matter over the space, it is
therefore necessary to give its realistic order of magnitude. Based
on the the relation (\ref{rho}), we evaluate roughly the order of
$\alpha$ in \textbf{TABLE \ref{table4}} in the four different
astronomical scales, i.e. Solar, Galaxy, Cluster of Galaxies and
Supercluster. In the range of Solar system, the measuring unit of
the distance between two planets is Astronomical Unit (AU). It is
well known that the nearest Mercury is about 0.4 AU and the most
remote Pluto is about 40 AU far from the sun. Here the average
orbital radius of Pluto is chosen as the value of the parameter $r$
in Solar system. The other three large astronomical scales are
estimated in the order of magnitude. If quintessence matter can be
treated as the dark energy model, its density $\rho_q$ can be
assumed as the critical density of the universe, namely,
\begin{equation}
\rho_q = \left\{
\begin{array}{c}
10^{-46}\ GeV^4\ \ \ \ \ (natural\ \ unit),\ \ \ \ \ \ \ \ \ \\
10^{-4}\ eV/cm^3\ \ \ \ (energy\ \ density\ \ unit),\\
10^{-29}\ g/cm^3\ \ \ \ (mass\ \ density\ \ unit).\ \ \ \\

\end{array}
\right.\label{rho11}
\end{equation}
Hence, substituting the density $\rho_q$ and the different scales
$r$ into relationship (\ref{rho}), we can get the order of magnitude
of $\alpha$ with different state parameter $\omega_q$, which is
shown in \textbf{TABLE \ref{table4}}. Apparently, the quintessence
matter makes more influence in the large scale. Comparing the
long-baseline radio interferometry experiment results \textbf{TABLE
\ref{table3}} with the theoretical expectation values \textbf{TABLE
\ref{table4}}, we can find that the case of $\omega_q = -2/3$
violates many orders of magnitude of $|\alpha|$ in solar. So the
case of $\omega_q = -2/3$ should be abandoned in solar system.

5 There are two easily confused points should be clarified in the
end. Firstly, the parameter $\alpha$ in Eq.(\ref{fg}) and
Eq.(\ref{metricq}) are the same as a matter of fact. The original
Kiselev's black hole solution is the general form of exact
spherically-symmetric solutions for the Einstein equations
describing black holes surrounded by the quintessential matter with
the energy momentum tensor yielding the additive and linear
conditions (\ref{add1}) and (\ref{theta}), i.e. Eqs.(13) --- (14) in
original paper \cite{Kiselev}. If we only consider the simplest
singular Schwarzschild case, the general solution to
Eq.(\ref{fequation}) should be in the form of
\begin{equation}\label{add1111}
e^{-\lambda} = 1 - \frac{r_g}{r} - \frac{\alpha}{r^{3\omega_q + 1}}.
\end{equation}
So the modified Schwarzschild black hole solution (\ref{metricq})
can be obtained without changing parameter $\alpha$. Mathematically,
if we adopt a singular field, not the mulriple fields, the first
term of Kiselev's solution should be
\begin{equation}\label{add2222}
\widetilde{g}_{00} =1 - \frac{r_g}{r} -
\left(\frac{r_0}{r}\right)^{3\omega_0 + 1}.
\end{equation}
 Comparing with the exact solutions (\ref{fg}) and (\ref{fbh}), we can get following equation
\begin{equation}\label{add3333}
\left(\frac{r_0}{r}\right)^{3\omega_0 + 1} =
\frac{\alpha}{r^{3\omega_q + 1}}.
\end{equation}
So it is clear that two relationships $r_0^{3\omega_0 + 1} = \alpha$
and $\omega_0 = \omega_q$ exist here. Hence, the parameters $\alpha$
in Eqs.(\ref{fg}) and (\ref{metricq}) are the same.

Secondly, the parameter $M$ is actually the mass of black hole (or
the center mass of gravitational field), not the mass of outside
quintessence matter. The form of quin-tessence field is denoted by
the energy momentum tensor yielding the conditions (8) and (9).
Since the dark energy almost has the ratio $\Omega_{de} \simeq 0.67
\pm 0.06$ in universe, the mass of quintessence $M_{q}$ is far more
than the mass of a stellar black hole $M$, i.e. $ M_{q} \gg M$. So
the quintessence density $\rho$ (\ref{rho}) does not depend largely
on the mass of black hole and $M$ can be assumed as a constant. This
assumption also can be found in the corresponding extensions of this
quintessence black hole solution \cite{quasinormal}, \cite{entropy},
\cite{geodetic}.

\begin{table*}[!h]
\caption{The theoretical magnitude order of parameter $\alpha$ in
several astronomical scales} \label{table4}
\begin{tabular}{lllll}
     \hline\noalign{\smallskip}
           astronomical scales & Solar (40AU) & Milky Way ($10^4$Pc)  & Cluster of Galaxies (1MPc)&Supercluster (10MPc)\\
     \noalign{\smallskip}\hline\noalign{\smallskip}
    $\omega_q = -1/3$ &$7.16\times10^{-13}\ Kg \centerdot m^{-1}$&$1.90\times10^{9}\ \ Kg \centerdot m^{-1}$&$1.90\times10^{13}\ Kg \centerdot m^{-1}$&$1.90\times10^{14}\ Kg \centerdot m^{-1}$\\
    $\omega_q = -2/3$&$5.98\times10^{-23}\ Kg \centerdot m^{-2}$&$3.08\times10^{-12} Kg \centerdot m^{-2}$&$3.08\times10^{-10} Kg \centerdot m^{-2}$&$3.08\times10^{-11} Kg \centerdot m^{-2}$\\
    $\omega_q =-1$&\multicolumn{4}{c}{$6.67 \times10^{-33}\ Kg \centerdot m^{-3}$}\\
\noalign{\smallskip}\hline
\end{tabular}
\end{table*}

\section{acknowledgement} Project supported by the
National Basic Research Program of China (2003CB716300), National
Natural Science Foundation of China (No.10573003) and National
Natural Science Foundation of China (No.10573004).

\appendix{
\section{the case of $a =2, \omega_q = 0$}\label{appendixa}
When the state parameter $\omega_q = 0$, Eq. (\ref{traqu2}) reduces
to
\begin{equation}\label{case1}
    \frac{d^2 u}{d \varphi^2} + u = \left(3 + \frac{3\alpha}{2M}\right) u^2.
\end{equation}
Because of the small $u$, we first ignore the 2th order small
quantity $3u^2$. The zeroth order approximate solution is
\begin{equation}\label{u01}
    u = u_0 \cos\varphi,
\end{equation}
where $u_0 = GM/R$ and $R$ is the solar radius. This is a straight
line normal to pole axis. Substituting Eq. (\ref{u01}) into the
right hand side of Eq. (\ref{case1}), we can get
\begin{equation}\label{u02}
    \frac{d^2 u}{d \varphi^2} + u = \left(3 + \frac{3\alpha}{2M}\right) u_0^2\cos^2\varphi.
\end{equation}
 This equation has a particular solution,
 \begin{equation}\label{particular1}
    u = \left(1 + \frac{\alpha}{2M}\right)u_0^2 (1 +
    \sin^2\varphi).
\end{equation}
So the first order approximate solution to Eq. (\ref{case1}) is
\begin{equation}\label{solutioncase1}
    u = u_0 \cos\varphi + \left(1 + \frac{\alpha}{2M}\right)u_0^2 (1 +
    \sin^2\varphi).
\end{equation}
The azimuth angle of null order approximate solutions (\ref{u01})
are $\pm\pi/2$ in very far place $u = 0$. However, the azimuth
angles of the first order one (\ref{solutioncase1}) are $\pm(\pi/2 +
\beta)$ in $u = 0$. The deflection angle $\beta$ is a small quantity
which satisfies
\begin{equation}\label{deflectionangle}
 -u_0 \sin\beta + \left(1 + \frac{\alpha}{2M}\right)u_0^2 (1 +
    \cos^2\beta) = 0.
\end{equation}
We use the Taylor expansions of sine and cosine functions and keep
the basic term. The deflection angle is written formally as
\begin{equation}\label{cae1deflecangle}
   \delta\varphi =2\beta = 4\left(1 + \frac{\alpha}{2 M}\right)\frac{GM}{R}.
\end{equation}
Comparing with the result of pure Schwarzschild space \cite{book},
the additional term containing parameter $\alpha$ is the
contribution of quintessence matter. The case of $\omega_q = 0$
affects the light deflection only through the particular solution
(\ref{particular1}). But the zeroth order approximate (\ref{u01}) is
unchanged.
\section{the case of $a =1, \omega_q = -1/3$}\label{appendixb}
When $\omega_q = -1/3$, Eq. (\ref{traqu2}) reduces to
\begin{equation}\label{case2}
    \frac{d^2 u}{d \varphi^2} + (1 - \alpha) u = 3 u^2.
\end{equation}
The process of solving this equation is similar to the formal one.
The zeroth order approximate solution is
\begin{equation}\label{u02}
    u = u_0 \cos (\varphi\sqrt{1 - \alpha}).
\end{equation}
The particular solution is
\begin{equation}\label{par2}
    u = \frac{u_0^2}{1 - \alpha} \left(1 + \sin^2 (\varphi \sqrt{1 - \alpha})\right).
\end{equation}
The first order approximate solution is
\begin{equation}\label{s2}
   u = u_0 \cos(\varphi\sqrt{1 - \alpha}) + \frac{u_0^2}{1 -
    \alpha}\left(1 + \sin^2(\varphi\sqrt{1 - \alpha})\right).
\end{equation}
So the deflection angle is
\begin{equation}\label{angle2}
 \delta\varphi = \frac{4}{(1-\alpha)^{3/2}}\frac{GM}{R}.
\end{equation}
The deflection behavior of the case $\omega_q = -1/3$ is different
from the case of $\omega_q = 0$. On the contrary, the quintessence
matter effects the photons deflection through the zero order
solution (\ref{u02}). So the later particular solution and first
order approximate solution vary accordingly.
\section{the case of $a =0, \omega_q = -2/3$}\label{appendixc}
When $\omega_q = -2/3$, Eq. (\ref{traqu2}) reduces to
\begin{equation}\label{case3}
    \frac{d^2 u}{d\varphi^2} + u - \frac{\alpha M}{2} = 3 u^2.
\end{equation}
In this case, the additional term $\alpha  M/2$ changes the form of
the zeroth order approximate solution. In order to simplify
calculation, we use a transformation
\begin{equation}\label{trancase3}
    u = y + \frac{1 - \sqrt{1-6\alpha M}}{6}.
\end{equation}
So the Eq. (\ref{case3}) can be rewritten as a usual orbital
equation
\begin{equation}\label{case33}
    \frac{d^2 y}{d \varphi^2} + \sqrt{1 - 6\alpha M} y= 3 y^2.
\end{equation}
Its first order approximate solution is
\begin{eqnarray}\label{casesolution}
y = y_0 \cos [\varphi(1 - 6 \alpha M)^{1/4}] + \frac{y_0^2}{\sqrt{1
- 6 \alpha M}} \bigg{\{} 1
     + \sin^2 [ \varphi (1 - 6 \alpha M)^{1/4}] \bigg{\}}.
\end{eqnarray}
Here we only consider the small contribution of quintessence matter
in the large space background. Hence, when $u \longrightarrow 0$,
the limit of $y$ is also small. So the deflection angle is
\begin{equation}\label{deflangle}
    \delta\varphi = \frac{4 y_0}{(1 - 6 \alpha M)^{3/4}},
\end{equation}
where
\begin{equation}\label{y0}
    y_0 = u_0 + \frac{\sqrt{1 - 6 \alpha M} - 1}{6}.
\end{equation}
\section{the case of $a =-1, \omega_q = -1$}\label{appendixd}
The additional term vanishes for the factor $\omega_q + 1 = 0$.
Hence, the influence of quintessence can be ignored. This point is
also justified by potential analysis. Since the exponential of
additional term is negative $3\omega_q + 2 < 0$, $u^a$ should not be
considered as a small quantity any more. The successive
approximation is failing here. However, the photon's effective
potential can be rewritten as
\begin{equation}\label{case4}
    \frac{1}{B^2(r)} = r^{-2} \left(1 - \frac{2 M}{r} - r^2 \alpha\right).
\end{equation}
The photons' trajectory equation (\ref{trajectoryequation}) of
$\omega_q = -1$ is
\begin{equation}\label{angularmo}
   \frac{d r}{d \varphi} = \pm r^2 \left[\frac{1}{b^2} - \frac{1}{B^2(r)}
   \right]^{1/2},
\end{equation}
where $b$ is the impact parameter and the $\pm$ denotes to
increasing or decreasing $r$. According to the condition of
perihelion position $r_0$, we can get
\begin{equation}\label{perihelion}
    \frac{d r}{d \varphi}\bigg|_{r = r_0} = \frac{1}{b^2} -
    \frac{1}{B^2(r_0)} = 0.
\end{equation}
So the impact parameter $b$ can be expressed by perihelion position
$r_0$. Submitting $b$ into Eq. (\ref{angularmo}), the quintessence
matter can be dropped out,
\begin{equation}\label{resultcase4}
    \frac{d r}{d \varphi} = \pm r^2 \left[\frac{1}{r_0}\left(1 - \frac{2 M}{r_0}\right) - \frac{1}{r^2}\left(1 - \frac{2
    M}{r}\right)\right]^{1/2}.
\end{equation}
Therefore, the quintessence matter has no influence on the light
deflection in this case, which is also justified by the articles
\cite{SdS}.}

\end{document}